\documentclass[aps,preprint]{revtex4}
\usepackage{graphicx}

\def \ba {\begin{eqnarray}}
\def \ea {\end{eqnarray}}
\def \bc {\begin{cases}}
\def \ec {\end{cases}}
\def \be {\begin{equation}}
\def \ee {\end{equation}}
\def \barr {\begin{array}}
\def \earr {\end{array}}
\def \bi {\begin{itemize}}
\def \ei {\end{itemize}}

\def \ra {\rightarrow}
\def \da {\downarrow}
\def \ua {\uparrow}

\newcommand{\llangle}{\langle\!\langle}
\newcommand{\rrangle}{\rangle\!\rangle}

\begin{document}

\title{Numerical renormalization group study
of the symmetric Anderson-Holstein model: phonon and electron
spectral functions}
\author{Gun Sang Jeon\footnote{Present address: The Pennsylvania State University,
Department of Physics, University Park, PA 16802.}, Tae-Ho Park,
and Han-Yong Choi\footnote{To whom the correspondences should be
addressed. e-mail: hychoi@skku.ac.kr.}}

\affiliation{Department of Physics, BK21 Physics Research
Division, and Center for Nanotubes and Nanoscale Composites, Sung
Kyun Kwan University, Suwon 440-746, Korea.}

%

\begin{abstract}

We study the symmetric Anderson-Holstein (AH) model at zero
temperature with Wilson's numerical renormalization group (NRG)
technique to study the interplay between the electron-electron and
electron-phonon interactions. An improved method for calculating
the phonon propagator using the NRG technique is presented, which
turns out to be more accurate and reliable than the previous
works in that it calculates the phonon renormalization explicitly
and satisfies the boson sum rule better. The method is applied to
calculate the renormalized phonon propagators along with the
electron propagators as the onsite Coulomb repulsion $U$ and
electron-phonon coupling constant $g$ are varied. As $g$ is
increased, the phonon mode is successively renormalized, and for
$g \gtrsim g_{co}$ crosses over to the regime where the mode
splits into two components, one of which approaches back to the
bare frequency and the other develops into a soft mode. The
initial renormalization of the phonon mode, as $g$ is increased
from 0, depends on $U$ and the hybridization $\Delta$; it gets
softened (hardened) for $U \gtrsim (\lesssim) U_s (\Delta)$.
Correlated with the emergence of the soft mode is the central
peak of the electron spectral function severely suppressed. These
NRG calculations will be compared with the standard Green's
function results for the weak coupling regime to understand the
phonon renormalization and soft mode.

\end{abstract}


\maketitle

\section{introduction}

An important problem in interacting many particle systems is
understanding the interplay between the electron-electron and
electron-phonon interactions. Such interplay will be important to
the problems such as the effects of phonons in the strongly
correlated electron systems and the effects of electron
correlations in the electron-phonon coupled systems. They include
the cuprates, organic superconductors, Kondo systems, fullerenes,
and the transition metal oxides, among others. Although some
perturbative approaches have been proposed to understand the
interplay \cite{millis1}, there has not been much progress in this
field partially because there are no reliable calculation schemes
applicable in strong coupling regime where standard perturbation
methods break down. Some non-perturbative techniques have been
developed for strongly correlated impurity model in the context
of the Kondo problem, but these techniques can not be applied to
the more interesting correlated models on a lattice. With the
recent advent of the dynamical mean-field theory (DMFT), however,
those techniques can be and have been successfully applied to a
wide class of lattice models. Within the DMFT, a lattice model is
mapped onto an effective impurity model \cite{metzner,georges}.
Among the various methods applied to solve the effective impurity
model, Wilson's numerical renormalization group (NRG) is
particularly powerful in that it is non-perturbative in nature so
that it can cover the whole parameter space, and it can be
applied to both zero and finite temperatures and to both the
dynamical and static properties
\cite{kristina1,kristina2,wilson1,bulla1}. The NRG+DMFT technique
has been successfully applied to the half-filled Hubbard model to
study the metal-insulator transition at both $T=0$ and finite
temperatures \cite{bulla2,bulla3}.

The prototype model for describing the interplay between
electrons and phonons is the Hubbard-Holstein model and the
Anderson-Holstein (AH) model which is the impurity version of the
Hubbard-Holstein model. Recently, the AH model was studied at
$T=0$ with the NRG technique by Hewson and Meyer \cite{hewson}.
The Holstein model was also studied with the DMFT combined with
the NRG technique \cite{meyer}. These works reported electron and
phonon spectral functions among others. In the present work, we
focus on the phonon propagator and report an improved method to
calculate the phonon spectral function using the NRG technique for
the AH model at zero temperature. This is the necessary first step
for applying the DMFT to the Hubbard-Holstein model. In section
II, we present the improved method for calculating the phonon
spectral function, which turns out to be more accurate and
reliable than the previous methods in that it calculates the
phonon renormalization explicitly and satisfies the sum rule
better \cite{hewson}. We will also present the standard Green's
function analysis for the AH model, which provides a useful
reference to understand the NRG results for the weak-coupling
regime. Then, in section III, the results of these NRG
calculations will be presented for the phonon and electron
spectral functions, and they will be compared with the standard
Green's function results to understand the phonon renormalization
and soft mode. Finally, section IV is for the summary and
concluding remarks.

\section{formulation}

The AH model is defined by
 \ba
 \label{ah1}
{\cal H}_{AH} &=& \sum_{k,\sigma} \epsilon_k c_{k\sigma}^\dag
c_{k\sigma} +\epsilon_f \sum_{\sigma} f_{\sigma}^\dag f_{\sigma}
+U f_{\ua}^\dag f_{\ua} f_{\da}^\dag f_{\da} +\omega_0 \left(
a^\dag a +{\frac12} \right)
 \\ \nonumber
&+& \sum_{k,\sigma} V_k \left( f_{\sigma}^\dag c_{k\sigma}
+c_{k\sigma}^\dag f_{\sigma} \right) + g \left( a^\dag +a \right)
\left( \sum_\sigma f_\sigma^\dag f_\sigma -1 \right),
 \ea
where $\sigma$ is the spin index, $\epsilon_f$ the impurity level
energy, and $V_k$ is the hybridization matrix element of the
impurity $f_\sigma $ electron with the conduction electrons of
the host metal described by the operators $c_{k\sigma}$. The $f$
electron has the onsite Coulomb repulsion $U$, and is linearly
coupled with the Einstein phonon of frequency $\omega_0$ with the
coupling constant $g$ at the impurity site. The coupling is taken
to be proportional to $(\sum_\sigma f_\sigma^\dag f_\sigma -1)$,
where the factor of $-1$ renders the problem particle-hole
symmetric for the symmetric AH model, that is, the electron
spectral function which will be discussed below is an even
function of $\omega$. The AH model of Eq.\ (\ref{ah1}) can be
rewritten as
 \ba
{\cal H}_{AH} &=& \left(\epsilon_f +\frac{U}{2} \right)
\sum_{\sigma} f_{\sigma}^\dag f_{\sigma} +\frac{U}{2} \left(
\sum_\sigma f_\sigma^\dag f_\sigma -1 \right)^2 +\omega_0 a^\dag a
 \\ \nonumber
&+& \sum_{k,\sigma} \epsilon_k c_{k\sigma}^\dag c_{k\sigma} +
\sum_{k,\sigma} V_k \left( f_{\sigma}^\dag c_{k\sigma}
+c_{k\sigma}^\dag f_{\sigma} \right) + g \left( a^\dag +a \right)
\left( \sum_\sigma f_\sigma^\dag f_\sigma -1 \right),
 \ea
disregarding constant terms. We will focus on the symmetric case
of $ \epsilon_f +U/2 =0$ in the present work for simplicity.
Asymmetric cases are also straightforward.

The electron and phonon Green's functions on the impurity site
are defined, respectively, by
 \ba
G_\sigma (\omega) = \llangle f_\sigma ,f_\sigma^\dag
\rrangle_\omega , ~~d(\omega) = \llangle a,a^\dag \rrangle_\omega,
 \label{def}
 \ea
where the correlator $\llangle ~~ \rrangle$ is defined as
 \ba
\llangle {\cal O}_1,{\cal O}_2 \rrangle_\omega &=&
\int_{-\infty}^{\infty} dt\ e^{i\omega t}\ \llangle {\cal
O}_1,{\cal O}_2 \rrangle_t ,
 \\ \nonumber
\llangle {\cal O}_1,{\cal O}_2 \rrangle_t &=& -i \theta(t) \left<
[{\cal O}_1 (t),{\cal O}_2 (0)]_\zeta \right>=-i \theta(t) \left<
[{\cal O}_1 (0),{\cal O}_2 (-t)]_\zeta \right>.
 \ea
Here, $\theta$ is the step function, $\left<~~\right>$ the
thermodynamic average, and $\zeta = +$ if both ${\cal O}_1$ and
${\cal O}_2$ are fermion operators and $\zeta = -$ otherwise. The
equation of motion (EOM) of the correlator is given by
 \ba
 \label{eom}
\omega \llangle {\cal O}_1,{\cal O}_2 \rrangle_\omega &=& \left< [
{\cal O}_1,{\cal O}_2 ]_\zeta \right> + \llangle [{\cal
O}_1,{\cal H} ]_- , {\cal O}_2 \rrangle_\omega ,
 \\ \nonumber
&=& \left< [ {\cal O}_1,{\cal O}_2 ]_\zeta \right> - \llangle
{\cal O}_1 , [{\cal O}_2,{\cal H} ]_- \rrangle_\omega .
 \ea
The $f$-electron Green's function can be written using the EOM
\cite{bulla-jpc98,hewson}
 \ba
G_{\sigma}^{-1} (\omega) = \omega-\epsilon_f -\bar\Delta(\omega)
-\Sigma_\sigma^U (\omega)-\Sigma_\sigma^g (\omega),
 \ea
where the self-energy corrections are given by
 \ba
 \label{selfs}
\bar\Delta(\omega) = \sum_k \frac{V_k^2}{\omega-\epsilon_k} ,~~
\Sigma_\sigma^U (\omega) = U \frac{\llangle f_\sigma
f_{-\sigma}^\dag f_{-\sigma} , f_\sigma^\dag \rrangle_\omega
}{\llangle f_\sigma ,f_\sigma^\dag \rrangle_\omega},~~
\Sigma_\sigma^g (\omega) = g\frac{\llangle f_\sigma (a^\dag +a )
, f_\sigma^\dag \rrangle_\omega }{\llangle f_\sigma ,f_\sigma^\dag
\rrangle_\omega}.
 \ea
This form of defining $G_\sigma (\omega)$ was proposed and its
advantages over the direct NRG calculation of $\llangle f,f^\dag
\rrangle_\omega $ were discussed by Bulla {\it et al.} and Hewson
and Meyer \cite{bulla-jpc98,hewson}.

Now, we consider the phonon propagator $d(\omega)$ defined by Eq.\
(\ref{def}). There are several ways to calculate the phonon
propagator: (a) We may calculate $d(\omega)=\llangle a,a^\dag
\rrangle_\omega$ directly by NRG, or, (b) calculate
 \ba
d^{-1}(\omega) &=& d_0^{-1} (\omega) -g\frac{\llangle n_f ,a^\dag
\rrangle_\omega}{\llangle a,a^\dag \rrangle_\omega} ,\label{defb2}
 \ea
where
 \ba
d_0^{-1}(\omega) = \omega-\omega_0 , ~~ n_f=\sum_\sigma
f_\sigma^\dag f_\sigma -1 ,
 \ea
or, (c) calculate
 \ba
d(\omega) &=& d_0(\omega) +[g d_0(\omega)]^2 \llangle n_f , n_f
\rrangle_\omega .
 \label{def3}
 \ea
Eqs.\ (\ref{defb2}) and (\ref{def3}) may be derived using the EOM
as follows. From the Eq.\ (\ref{eom}), we have
 \ba
(\omega-\omega_0) \llangle a,a^\dag \rrangle_\omega = 1 +g
\llangle n_f, a^\dag \rrangle_\omega .
 \label{defb2-a}
 \ea
Eq.\ (\ref{defb2-a}) can be rearranged, by dividing both sides by
$\llangle a,a^\dag \rrangle_\omega$, to Eq.\ (\ref{defb2}). Again
from the EOM, we also obtain
 \ba
\omega \llangle n_f , a^\dag \rrangle_\omega = \omega_0 \llangle
n_f , a^\dag \rrangle_\omega + g \llangle n_f , n_f
\rrangle_\omega .
 \ea
Substituting this into Eq.\ (\ref{defb2-a}) yields Eq.\
(\ref{def3}).

The advantage of using Eq.\ (\ref{defb2}) is that it is written in
the form of Dyson equation and shows the renormalization
explicitly. The form of Eq.\ (\ref{def3}) was employed previously
\cite{hewson,meyer}. It will be extremely difficult, however, to
obtain the renormalized spectrum near $\omega \approx \omega_0$
from Eq.\ (\ref{def3}) because of the strong singularity of the
$[d_0(\omega)]^2$. We therefore use Eq.\ (\ref{defb2}) in the
present analysis. This argument is made more quantitative by the
sum rule. The important point is that the $d(\omega)$ must satisfy
the boson sum rule:
 \ba
-\frac{1}{\pi} \int_{-\infty}^{\infty}d\omega\ Im\ d(\omega) =1.
 \label{sum-rule}
 \ea
The three ways of calculating the phonon propagator above will be
discussed in the following section. It will be shown that the
method (b) satisfies the sum rule of Eq.\ (\ref{sum-rule}) much
better than the other methods.

Alternatively, one may use the phonon Green's function defined by
 \ba
D(\omega) = \llangle a+a^\dag, a+a^\dag \rrangle_\omega.
 \ea
We then have
 \ba
D^{-1}(\omega) = D_0^{-1}(\omega) -g\frac{\llangle n_f , a+a^\dag
\rrangle_\omega}{\llangle a+a^\dag, a+a^\dag \rrangle_\omega},
 \\ \nonumber
D(\omega) = D_0 (\omega) +[g D_0 (\omega)]^2 \llangle n_f , n_f
\rrangle_\omega ,
 \ea
which correspond to Eqs. (\ref{defb2}) and (\ref{def3}),
respectively, where $D_0 (\omega) = 2\omega_0/(\omega^2
-\omega_0^2 ) $. There are essentially no differences between the
phonon Green's functions $d(\omega)$ and $D(\omega)$. The
advantage of $d(\omega)$ over $D(\omega)$ is that there exists the
useful sum rule of Eq.\ (\ref{sum-rule}) the $d(\omega)$
satisfies, while the sum rule the $D(\omega)$ satisfies
 \ba
- \frac{1}{\pi} \int_{-\infty}^{\infty} d\omega Im D(\omega)
n(\omega) = \left< (a^\dag +a)(a^\dag +a) \right> ,
 \ea
where $n(\omega)$ is the Bose function, is less useful.

\begin{figure}
\label{fig1}
\includegraphics[scale=0.4]{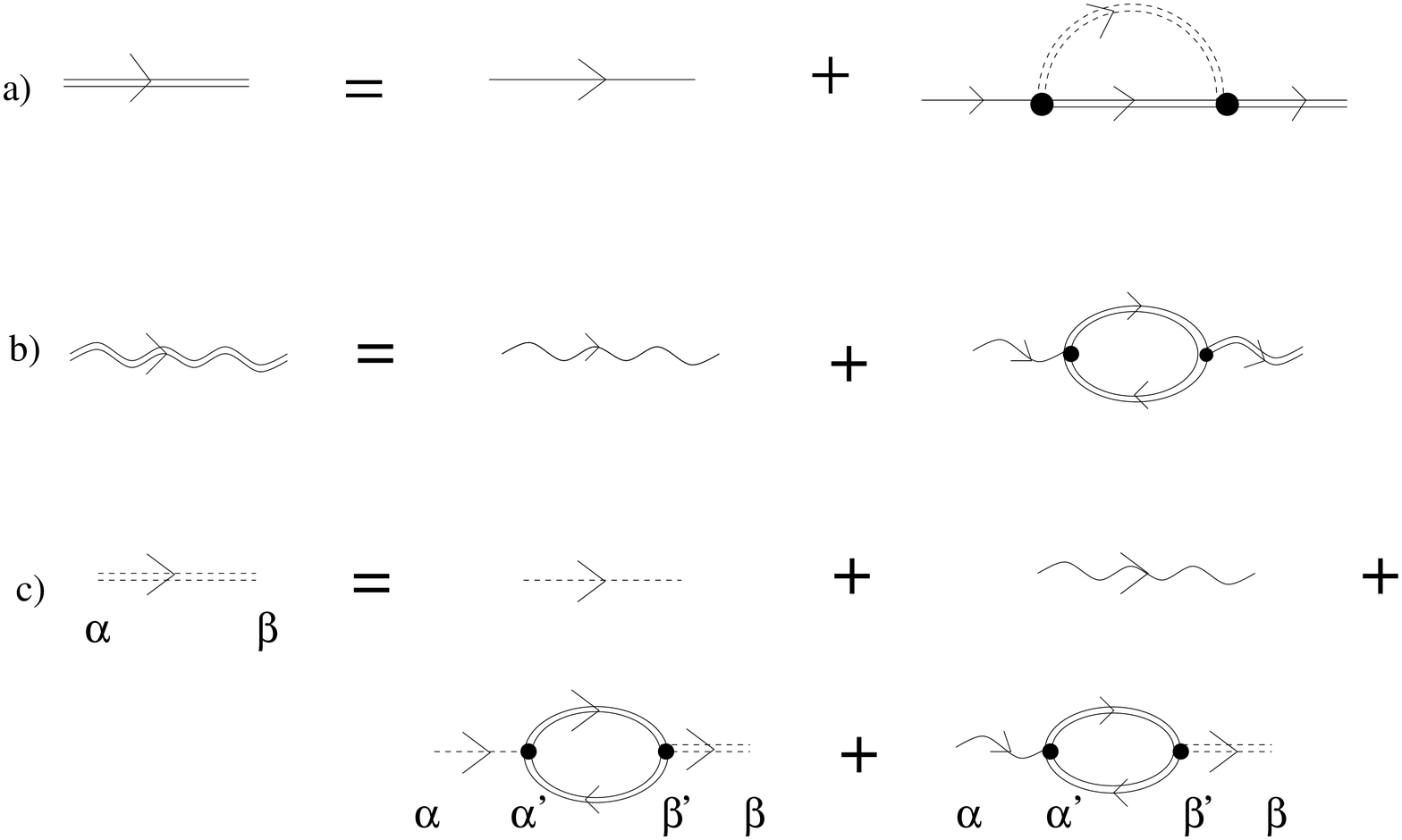}
\caption{The Feynman diagram for the Anderson-Holstein model. The
single (double) solid and wiggle lines represent, respectively,
the bare (renormalized) electron and phonon propagators. The
single dashed line refers to the Hubbard interaction, and the
double dashed line to the effective interaction between two
electrons of spin $\alpha$ and $\beta$.}
\end{figure}

Before we present the detailed NRG results for the AH model, we
consider the standard Green's function analyses of the model for
the phonon and electron propagators. It will provide a useful
reference to understand the NRG results for the weak-coupling
regime. Within the perturbative Green's function scheme, the
renormalized electron and phonon propagators for $U=0$ are, as
shown in Fig.\ 1(a) and (b), given by
 \ba
G^{-1}(\omega) = G_0^{-1}(\omega) -\Sigma(\omega), ~~
D^{-1}(\omega) = D_0^{-1}(\omega) -g^2 \chi(\omega),
 \ea
where $G_0$ is the bare electron Green's function given by
 \ba
 \label{g0}
G_0 (\omega) = \frac{1}{\omega+i\Delta},~~ \Delta = \sum_k \pi
|V_k |^2 \delta(\omega-\epsilon_k ),
 \ea
and the self-energy $\Sigma$ and the polarization $\chi$ are
given by
 \ba
\Sigma(\omega) &=& \frac{1}{\pi^2} \int_{-\infty}^{\infty}
d\epsilon \int_{-\infty}^{\infty} d\Omega
\frac{f(\epsilon)+n(-\Omega)} {\epsilon+\Omega-\omega-i\delta} Im
G(\epsilon) Im V(\Omega),
 \\ \nonumber
\chi(\omega) &=& \frac{1}{\pi^2} \int_{-\infty}^{\infty} d\epsilon
\int_{-\infty}^{\infty} d\epsilon' \frac{f(\epsilon)-f(\epsilon')}
{\omega+\epsilon-\epsilon' +i\delta} Im G(\epsilon) Im
G(\epsilon').
 \ea
$\Delta$ of Eq.\ (\ref{g0}) is the minus imaginary part of
$\bar\Delta$ defined in Eq.\ (\ref{selfs}),
$\bar\Delta=-i\Delta$. $V$ is the effective interaction between
two electrons which is reduced to $V=g^2 D$ for $U=0$ as discussed
below. We consider the case where $\Delta(\omega)$ is a constant.
At $T=0$ and up to the 2nd order in $g$, we obtain
 \ba
 \label{chi}
\chi(\omega) &=& -\frac1\pi \int_0^\infty d\epsilon
\frac{\Delta}{\epsilon^2 +\Delta^2} \left(
\frac{1}{\epsilon+\omega+i\Delta}
+\frac{1}{\epsilon-\omega-i\Delta} \right)
 \nonumber \\
&=& \frac{2\Delta}{\pi\omega} \frac{1}{\omega+i 2\Delta}
\ln\left(1-i\frac{\omega}{\Delta} \right) .
 \ea
The real and imaginary parts are given, respectively, by
 \ba
 \label{chi12}
Re \chi(\omega) &=& \frac{\Delta}{\pi} \frac{1}{4\Delta^2
+\omega^2} \left[ \ln \left( 1+ \frac{\omega^2}{ \Delta^2} \right)
-\frac{4\Delta}{\omega} \tan^{-1} \frac\omega\Delta \right],
 \nonumber\\
Im \chi(\omega) &=& -\frac{2\Delta^2}{\pi\omega}
\frac{1}{4\Delta^2 +\omega^2} \left[ \ln \left( 1+
\frac{\omega^2}{ \Delta^2} \right) +\frac\omega\Delta \tan^{-1}
\frac\omega\Delta \right].
 \ea
For $\omega \ll \Delta$, it is reduced to
 \ba
 \label{chismall}
\chi(\omega) \approx -\frac{1}{\pi\Delta} \left( 1-\frac56
\frac{\omega^2}{\Delta^2} +i\frac{\omega}{\Delta} \right),
 \ea
which will be used in discussing the soft phonon mode in the
following section.

When $U\ne 0$, the renormalized phonon propagator is given by
 \ba
D^{-1} (\omega) = D_0^{-1} (\omega) - g^2
\frac{\chi(\omega)}{1-U\chi(\omega)/2} .
 \label{ren-d}
 \ea
This may be obtained as follows. The effective interaction
between the $f$-electrons $V_{\alpha\beta}$, where $\alpha$ and
$\beta$ are the spin indices, due to both electron-phonon and
electron-electron interactions is determined, as shown in Fig.
1(c), by
 \ba
V_{\alpha\beta} (i\omega) = V_{\alpha\beta}^0 +
\sum_{\alpha',\beta'} V_{\alpha\alpha'}^0 \chi_{\alpha'\beta'}
(i\omega ) V_{\beta'\beta} (i\omega ), \label{v-chi_ab}
 \ea
where
 \ba
V_{\alpha\alpha}^0 &\equiv& V^0 = g^2 D_0,~~ V_{\alpha,-\alpha}^0
= U +V^0.
 \ea
The solution to Eq.\ (\ref{v-chi_ab}) is given by
 \ba
V_{\uparrow\uparrow} &=& V_{\downarrow\downarrow}= \frac{V^0
+U(2V^0 +U) \chi_{\uparrow\uparrow}} {\left[1-(2V^0
+U)(\chi_{\uparrow\uparrow} +\chi_{\uparrow\downarrow}) \right]
\left[1+U(\chi_{\uparrow\uparrow} -\chi_{\uparrow\downarrow})
\right]},
\nonumber \\
V_{\uparrow\downarrow} &=& V_{\downarrow\uparrow}= \frac{V^0 +U
-U(2V^0 +U) \chi_{\uparrow\downarrow}} {\left[1-(2V^0
+U)(\chi_{\uparrow\uparrow} +\chi_{\uparrow\downarrow}) \right]
\left[1+U(\chi_{\uparrow\uparrow} -\chi_{\uparrow\downarrow})
\right]}.
 \ea
We put $ \chi_{\uparrow\uparrow} = \chi_{\downarrow\downarrow} =
\chi/2$ and $\chi_{\uparrow\downarrow} =
\chi_{\downarrow\uparrow} = 0$ to obtain
 \ba
V_{\ua\ua} = \frac{V^0 +U(V^0 +U/2) \chi} {\left[1-(V^0 +U/2)\chi
\right] \left[1+U\chi/2 \right]},
 \nonumber \\
V_{\ua\da} = \frac{V^0 +U} {\left[1-(V^0 +U/2)\chi \right]
\left[1+U\chi/2 \right]}.
 \ea
The renormalized electron-phonon interaction is then given by
\cite{mahan}
 \ba
V_{el-ph} &=& V_{\ua\ua} -\frac{U^2 \chi/2}{1-U^2\chi^2 /4} =
V_{\ua\da} -\frac{U}{1-U^2\chi^2 /4}
 \nonumber \\
&=& \frac{V^0}{[1-(2V^0+U)\chi/2][1-U\chi/2]}.
 \ea
We put $V^0 = g^2 D_0$ to obtain
 \ba
V_{el-ph} (\omega) = \frac{g^2}{\left[ 1-U\chi(\omega) /2
\right]^2 } D(\omega),
 \label{ren-ph}
 \ea
where $D(\omega)$ and $\chi(\omega)$ are given, respectively, by
Eqs.\ (\ref{ren-d}) and (\ref{chi}). We will use the renormalized
phonon propagator of Eq.\ (\ref{ren-d}) in the following section
to understand the phonon renormalization and emergence of the soft
phonon mode.

\section{spectral functions and sum rule}

The renormalized phonon Green's function $d(\omega)$ was obtained
by the method (b) of Eq.\ (\ref{defb2}) of the previous section by
calculating $\llangle n_f ,a^\dag \rrangle_\omega$ and $\llangle
a,a^\dag \rrangle_\omega$ directly with the NRG technique. Some
technical details of the NRG scheme can be found in Ref.\
\cite{meyer,bulla-jpc98}. Briefly, the conduction electron band
of the AH model, which was taken from $-1$ to 1, is
logarithmically discretized and mapped onto a semi-infinite
fictitious chain coupled to the impurity site at one end such
that the hopping amplitude along the chain falls off expontially
($\sim\Lambda^{-n/2}$). The many body eigen-states and energies
are calculated iteratively starting from the impurity site and
adding an extra site along the chain at each step. After a few
steps, the number of states has to be truncated to the lowest
$N_{st}$ ones. In the present work, we used the discretization
parameter $\Lambda=2.0 $, kept $N_{st}= 600$ states in each
iteration, and 100 fictitious sites along the chain. The lowest
40 phonon states are kept for the AH model with the bare phonon
frequency $\omega_0 = 0.05$. The energy unit is the half the
conduction bandwidth.

\begin{figure}
\label{fig2} \vspace{5cm}
\includegraphics[scale=0.5]{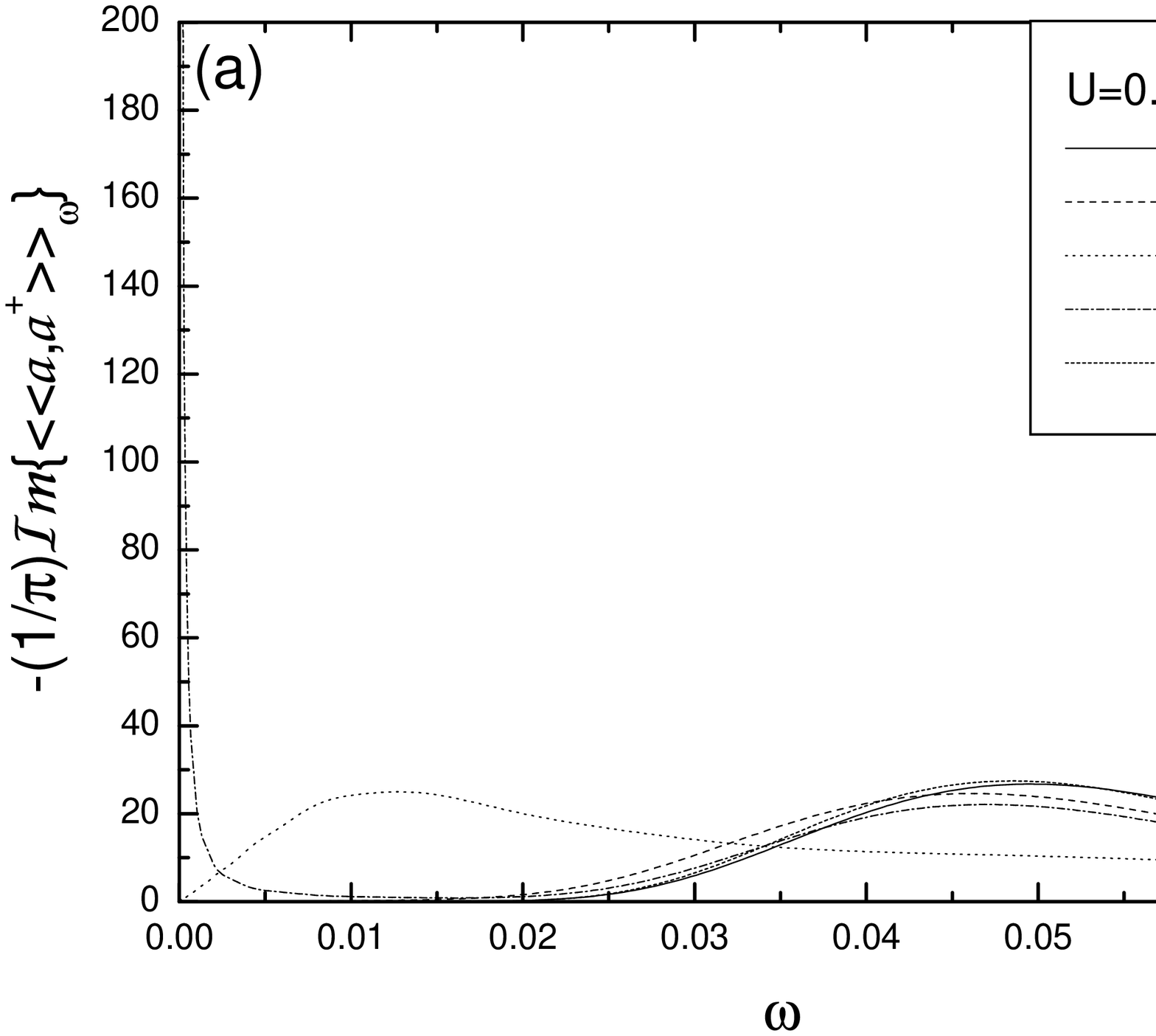}
\includegraphics[scale=0.5]{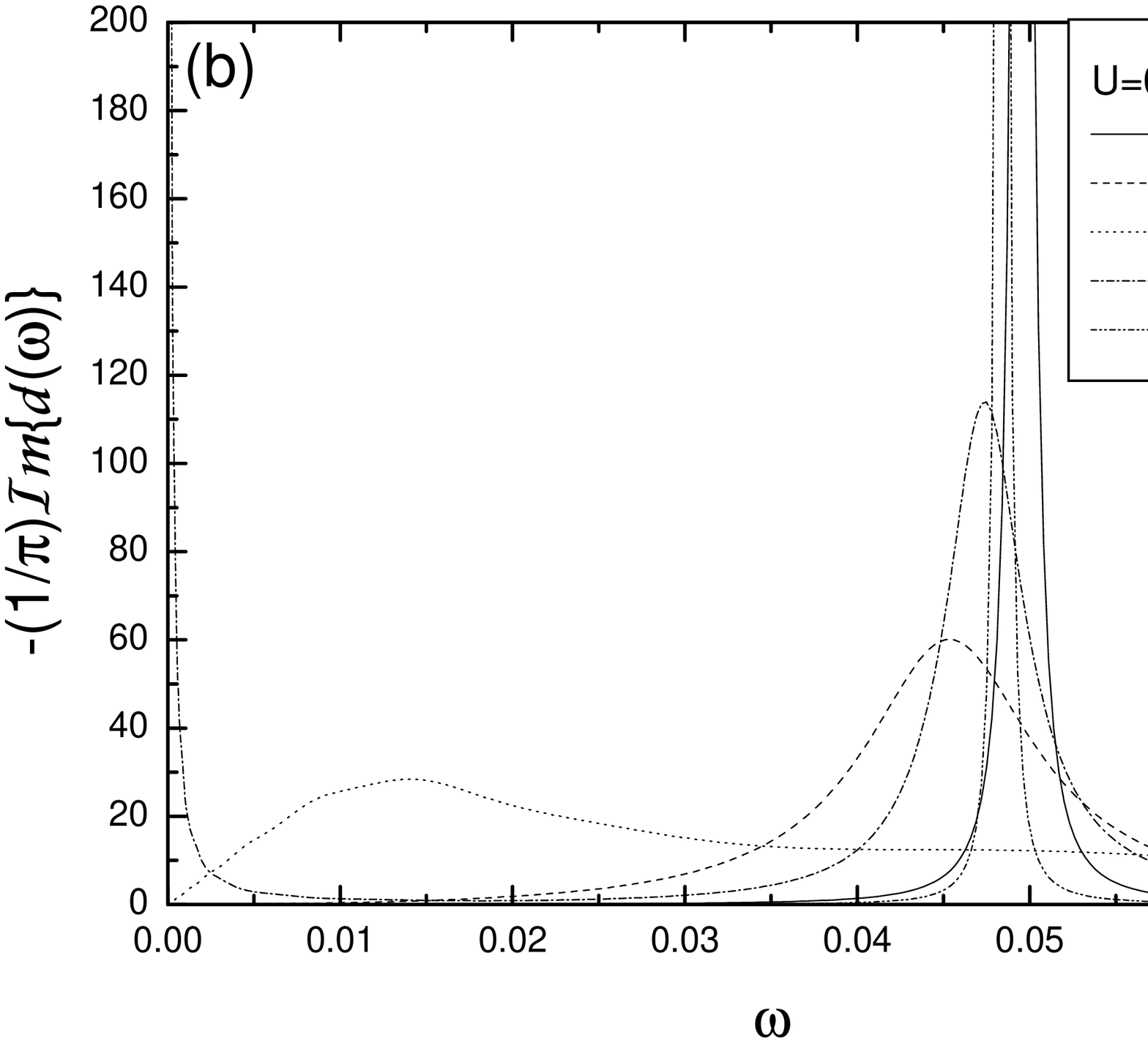}
\end{figure}

\begin{figure}
\vspace{5cm}
\includegraphics[scale=0.5]{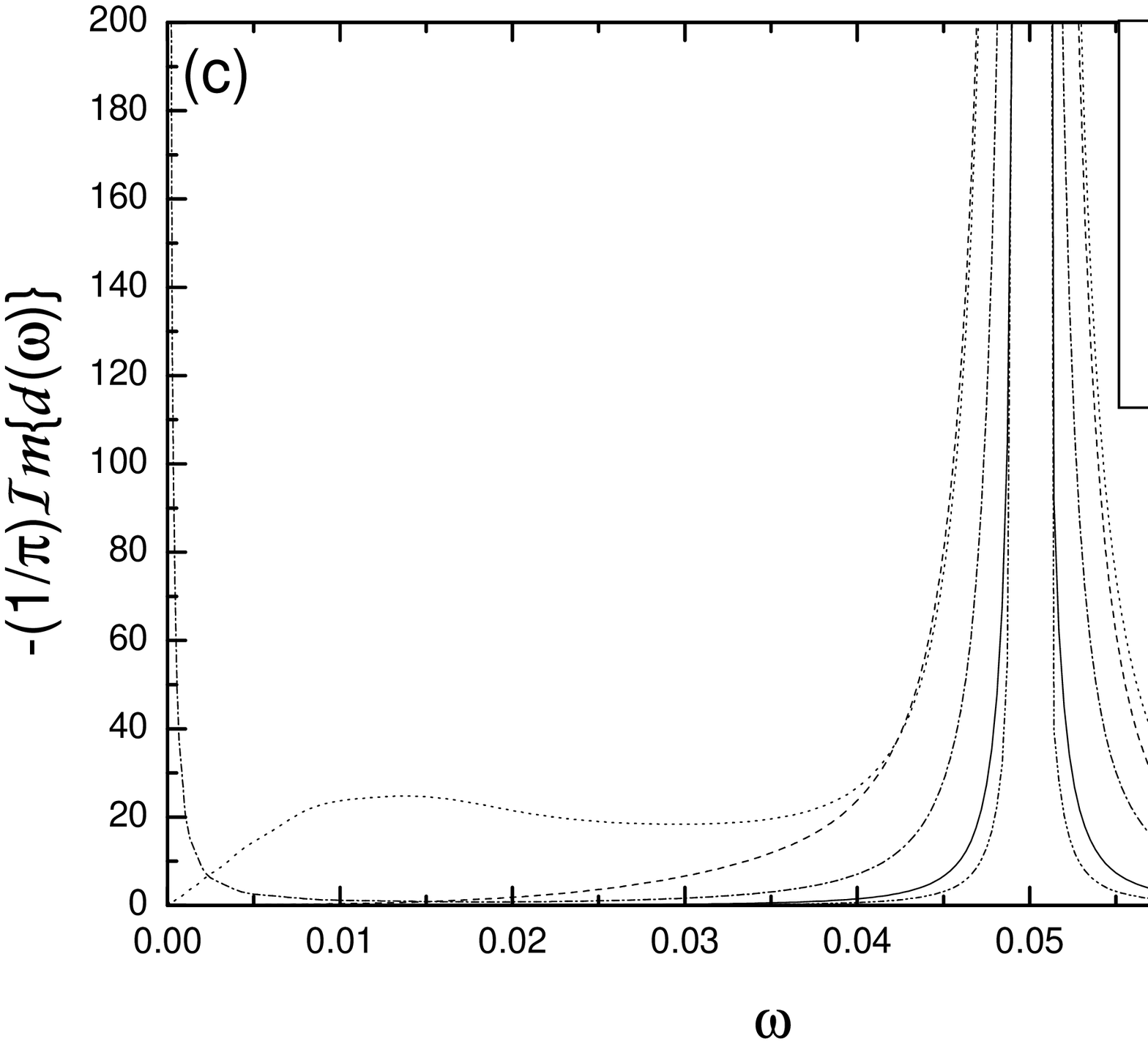}
\vspace{-5cm} \caption{The phonon spectral function
$-\frac{1}{\pi} Im \, d(\omega)$ vs.\ $\omega$ calculated by the
three methods for $U=0.1$ and $\Delta=\omega_0/\pi \approx 0.0159
$. Figure (a), (b), and (c) are obtained by the method (a) the
direct NRG calculation of $ \llangle a,a^\dag \rrangle_\omega$,
(b) the present calculation of Eq.\ (\ref{defb2}), and (c) Hewson
and Meyer of Eq. (\ref{def3}), respectively.}
\end{figure}

We now compare the three ways of calculating the $d(\omega)$ by
the methods (a), (b), and (c), given, respectively, by Eqs.\
(\ref{def}), (\ref{defb2}), and (\ref{def3}) explained in the
previous section. In Fig.\ 2, we show $-\frac{1}{\pi}Im\,
d(\omega)$ vs.\ $\omega$ calculated by the above three methods
for $U=0.1$ and $\Delta=\omega_0/\pi \approx 0.0159$, for the
electron-phonon coupling constants $g=$ 0.01, 0.03, 0.05, 0.07,
and 0.09. The NRG calculations yield the discrete frequencies;
the resulting spectral functions are discrete with the
$\delta$-peaks getting closer as $\omega\ra 0$. The
$\delta$-peaks are broadened via
 \ba
\delta(\omega-\omega_n ) \ra \frac{e^{-b^2 /4}}{b\, \omega_n
\sqrt{\pi}} \exp\left[-\frac{(\ln\omega-\ln\omega_n )^2}{b^2}
\right].
 \ea
This broadening function is a Gaussian on a logarithmic scale
with the width $b$. $b=0.4$ was used in the present study. Figures
(a), (b), and (c) are the results obtained by the methods (a),
(b), and (c), respectively. For the low frequency, they show the
same behavior, but for $\omega \approx \omega_0$, they show
deviations from each other. In the method (a) the feature around
the bare frequency $\omega_0$ is almost washed out, while in the
method (c) the peak is always tied to the bare frequency because
of the singularity of the $[d_0 (\omega)]^2$. In the method (b),
the phonon frequency is successively renormalized as one may
expect from the perturbation theory. As the electron-phonon
coupling constant $g$ is increased, the phonon mode softens until
the soft mode begins to emerge around $g \approx g_{co}$, which
will be discussed below. As $g$ is further increased above
$g_{co}$, the mode splits into two components, one of which
hardens back to the bare frequency and the other develops into
the soft mode as can be seen for $g=0.07$ and 0.09 of the figure
(b). For $g=0.09$, the soft mode can hardly be seen because the
width of the soft mode is very narrow in this energy scale.

In the table 1, we compare the integrals $-\frac{1}{\pi}
\int_{-\infty}^{\infty} d\omega Im \, d(\omega)$ from the method
(a) and (b) to check the sum rule of Eq.\ (\ref{sum-rule}). The
method (b) satisfies the sum rule better than (a), and the method
(c) doesn't even converge because of the strong singularity of
the $[d_0 (\omega)]^2$ in Eq.\ (\ref{def3}). Because of the
singularity, it will be extremely difficult to obtain the phonon
renormalization around $\omega \approx \omega_0$ from the method
(c). The Eq.\ (\ref{defb2}) of method (b) is written in the form
of Dyson equation where $(1/g) \llangle n_f ,a^\dag
\rrangle_\omega / \llangle a,a^\dag \rrangle_\omega$ is the
proper polarization, which gives the phonon renormalization as
$g$ is increased. As $g$ is further increased, the soft mode of
$\omega=0$ emerges, which may be understood in terms of the
Green's function perturbation scheme discussed below. The present
method (b) of Eq.\ (\ref{defb2}) calculated the phonon
renormalization explicitly and satisfies the sum rule better than
the other methods and, therefore, give more reliable and
satisfactory description of the phonon renormalization of the AH
model.

\begin{table}
\begin{tabular}[c]{|c|c|c|}
 \hline
  $g$  & Direct NRG & Improved method \\ \hline
  0.01 & 1.000046 & 0.999922 \\
  0.03 & 0.999212 & 0.999890 \\
  0.05 & 0.993977 & 0.999863 \\
  0.07 & 0.911200 & 0.991001 \\
  0.09 & 0.992165 & 0.999840 \\ \hline
\end{tabular}
\caption{The boson sum rule of Eq.\ (\ref{sum-rule}). The second
and third columns show, respectively, the integral $-\frac1\pi
\int_{-\infty}^{\infty} d\omega Im \, d(\omega)$ computed by the
direct NRG calculation of $\llangle a,a^\dag \rrangle_\omega $ and
the improved method of Eq.\ (\ref{defb2}). }
\end{table}

The renormalized phonon frequency can be found from $D^{-1}
(\omega) = 0$. To investigate the emergence of the soft phonon
mode, we substitute the expression Eq.\ (\ref{chismall}) of
$\chi(\omega)$ for $\omega \ll \Delta$ into Eq.\ (\ref{ren-d}) to
obtain
 \ba
\omega^2 \left[1- \frac{4\pi g^2 \omega_0 (10\pi\Delta -U)}
{3\Delta (2\pi\Delta +U)^3 }\right] - \omega_0 \left( \omega_0
-\frac{4g^2}{2\pi\Delta +U} \right) = 0.
 \ea
The cross-over value $g_{co} $ is then given by
 \ba
 \label{soft}
g_{co} = \frac12 \sqrt{\omega_0 \left(2\pi\Delta+ U \right)}
 \ea
which gives $g_{co} = 0.05$ for $U=0.1$ and $\Delta=\omega_0/\pi$.
We therefore expect the soft mode emerges for $g \gtrsim g_{co}$
which agrees fairly well with the NRG calculations of Fig.\ 2(b).

\begin{figure}
 \label{fig3}
\vspace{5cm}
\includegraphics[scale=0.5]{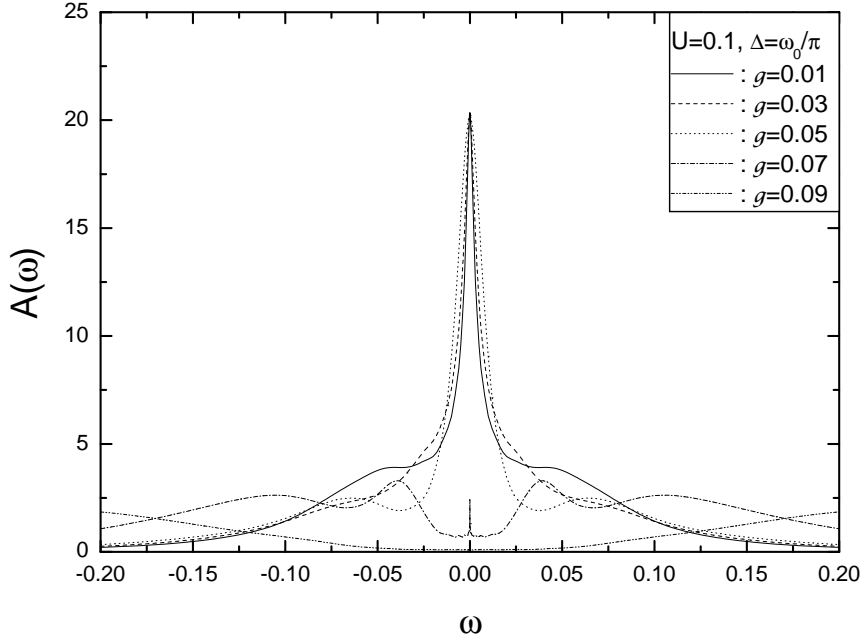}
\vspace{-5cm} \caption{The electron spectral function
$-\frac{1}{\pi} Im G(\omega)$ vs.\ $\omega$ for $U=0.1$ and
$\Delta=\omega_0/\pi$ corresponding to the figure 2. }
\end{figure}

In Fig.\ 3, we show the electron spectral function
$A(\omega)=-\frac{1}{\pi} Im G(\omega)$ vs.\ $\omega$ for the same
parameters as the Fig.\ 2. For small $g$ such that $\bar U
>\pi\Delta$, where $\bar U =U-2 E_p$ is the effective
onsite repulsion and $E_p = g^2/\omega_0$ is the polaron binding
energy, the system is in the Kondo regime. The peak at $\omega=0$
is then the Kondo resonance peak whose width is given by
 \ba
 \label{width}
2 T_K \simeq (\bar U\Delta)^{\frac{1}{2}} \, \exp(-\frac{\bar
U}{\pi\Delta}),
 \ea
and the height is
 \ba
 \label{height}
A(\omega=0) = \frac{1}{\pi\Delta},
 \ea
in agreement with the figure 3. The shoulders at $\omega \approx
\pm \bar U /2 \approx \pm 0.05$ correspond to the charge
fluctuations at the $f$ electron impurity site. As $g$ is
increased, $\bar U$ is decreased, and the width of the Kondo peak
increases as can be seen from the figure. The effective $\bar U$
becomes negative for $g>g_-=\sqrt{U\omega_0/2}$. $g_-=0.05$ for
$U=0.1$ and $\omega_0 =0.05$. The electron spectral function in
this case may be understood in terms of the small polaron picture
as discussed by Hewson and Meyer \cite{hewson}. The electron
Green's function for small hybridization limit of $\Delta \ra 0$
is given by
 \ba
 \label{smallpol}
G_\sigma(\omega)=  e^{-\lambda} \sum_{n=0}^{\infty}
\frac{\lambda^n}{ n!} \bigg[ \frac{ \left<
(1-n_{f,\sigma})(1-n_{f,-\sigma})\right>}
{\omega-\bar\epsilon_f-n\omega_0} +\frac{\left<
(1-n_{f,\sigma})n_{f,-\sigma}\right>}{ \omega-\bar\epsilon_f-\bar
U-n\omega_0}
 \nonumber\\
+ \frac{ \left<n_{f,\sigma}(1-n_{f,-\sigma})\right>}{
\omega-\bar\epsilon_f+n\omega_0} + \frac{ \left<
n_{f,\sigma}n_{f,-\sigma}\right>}{ \omega-\bar\epsilon_f-\bar
U+n\omega_0}\bigg],
 \ea
where $\lambda=g^2/\omega_0^2$, $ \left< n_{f,\sigma} \right>
=\left<f_\sigma^\dag f_\sigma \right>$, and $\bar{\epsilon}_f =
\epsilon_f +E_p$. Note that $\bar{\epsilon}_f = \epsilon_f +E_p$
instead of $\bar{\epsilon}_f = \epsilon_f -E_p$ because the
coupling between the $f$ electrons and the phonons is taken to be
proportional to $\left( \sum_\sigma f_\sigma^\dag f_\sigma -1
\right)$ instead of $\left( \sum_\sigma f_\sigma^\dag f_\sigma
\right)$. The set of delta functions at $\omega =\bar\epsilon_f
+n\omega_0,~\bar\epsilon_f-n\omega_0,~\bar\epsilon_f
+\bar{U}-n\omega_0$ and $\bar\epsilon_f +\bar{U}-n\omega_0$ for
integer $n$, determined by the denominators of Eq.\
(\ref{smallpol}), will be broadened for non-zero $\Delta$. The
most probable phonon occupation number is $\left< n
\right>=E_p/\omega_0 = \lambda$. For integer $n$'s, the peaks are
expected at
 \ba
 \label{polaronpeaks}
\omega &=&\bar\epsilon_f \pm n\omega_0 = \epsilon_f
+E_p,~\epsilon_f +E_p \pm \omega_0,\cdots,
 \nonumber\\
\omega &=&\bar\epsilon_f +\bar{U} \pm n\omega_0 = \epsilon_f
+U-E_p,~\epsilon_f +U-E_p \pm \omega_0,\cdots,
 \ea
which roughly tracks the four high energy peaks of the spectral
function $A(\omega)$ as $g$ is increased. For $g=0.07$ for
instance, $E_p=0.098$, and we expect the peaks at $\omega = \pm
0.048,~\pm0.098$, and so forth, which roughly agree with the
detailed NRG results shown in the figure 3. The central peak for
the negative $\bar U$ may be understood in terms of the Kondo-like
resonance of the negative $U$ model, which arises because of the
charge fluctuations, instead of the spin fluctuations in the Kondo
regime, between the zero and double occupation states at the
impurity site. When $g$ is further increased above $g_{co}$, the
soft phonon mode emerges. The effective $| \bar U|$, which equals
to $2g^2/\omega_0 -U$ for the Einstein phonon may be generalized
to
 \ba
|\bar U|= 2g^2 \left< \frac1\omega\right> -U, ~~{\rm where}~
\left< \frac1\omega\right> = -\int_{-\infty}^{\infty} d\omega
\frac1\pi Im\, d(\omega) \frac1\omega .
 \ea
When the soft mode emerges, the $\bar U$ can be very large and the
width of the electron central peak given by Eq.\ (\ref{width})
becomes very narrow. The central peak can hardly be seen for
$g=0.07$ and 0.09 of the Fig.\ 3 in this frequency scale.

\begin{figure}
 \label{fig4}
\vspace{5cm}
\includegraphics[scale=0.5]{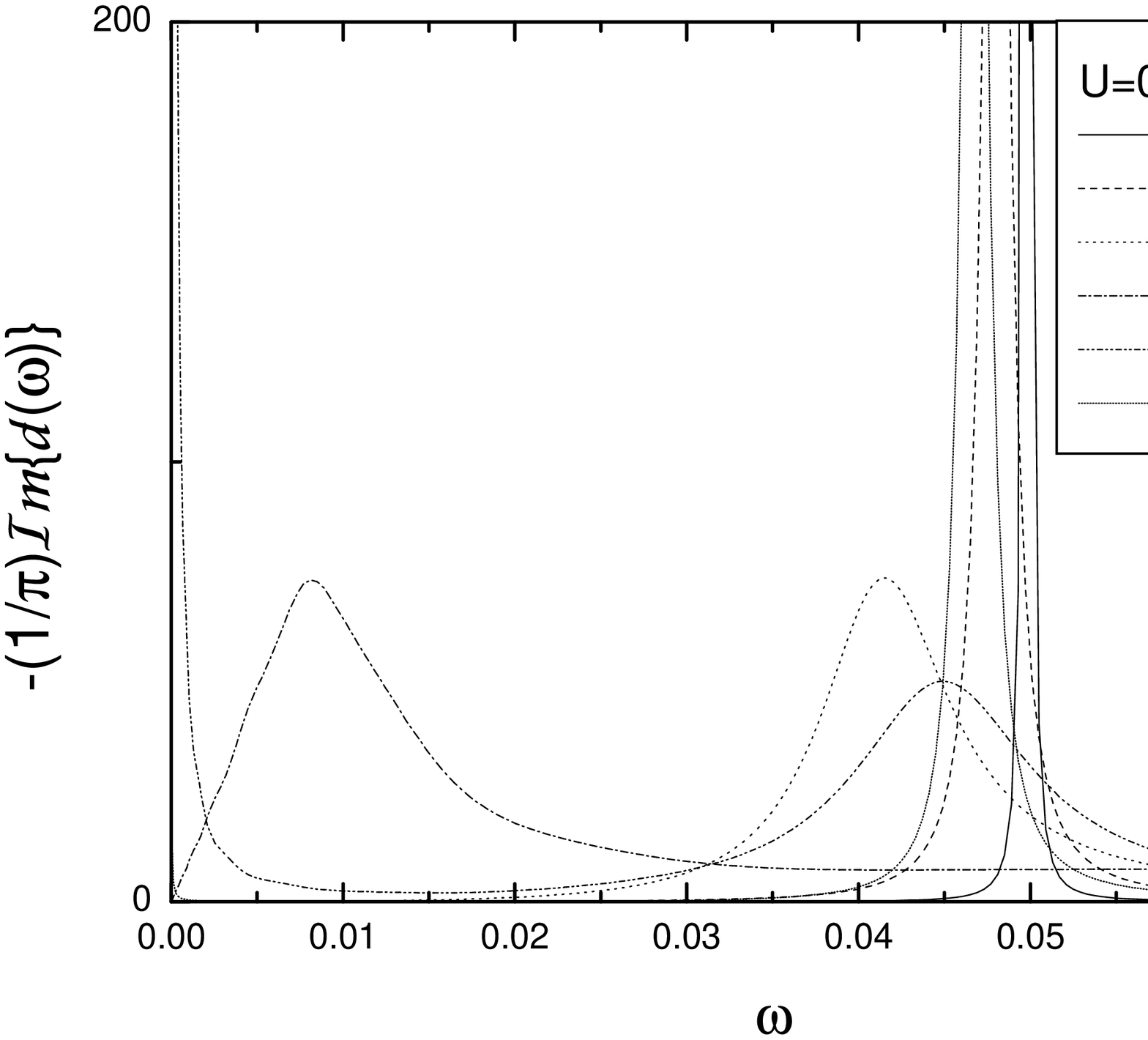}
\includegraphics[scale=0.5]{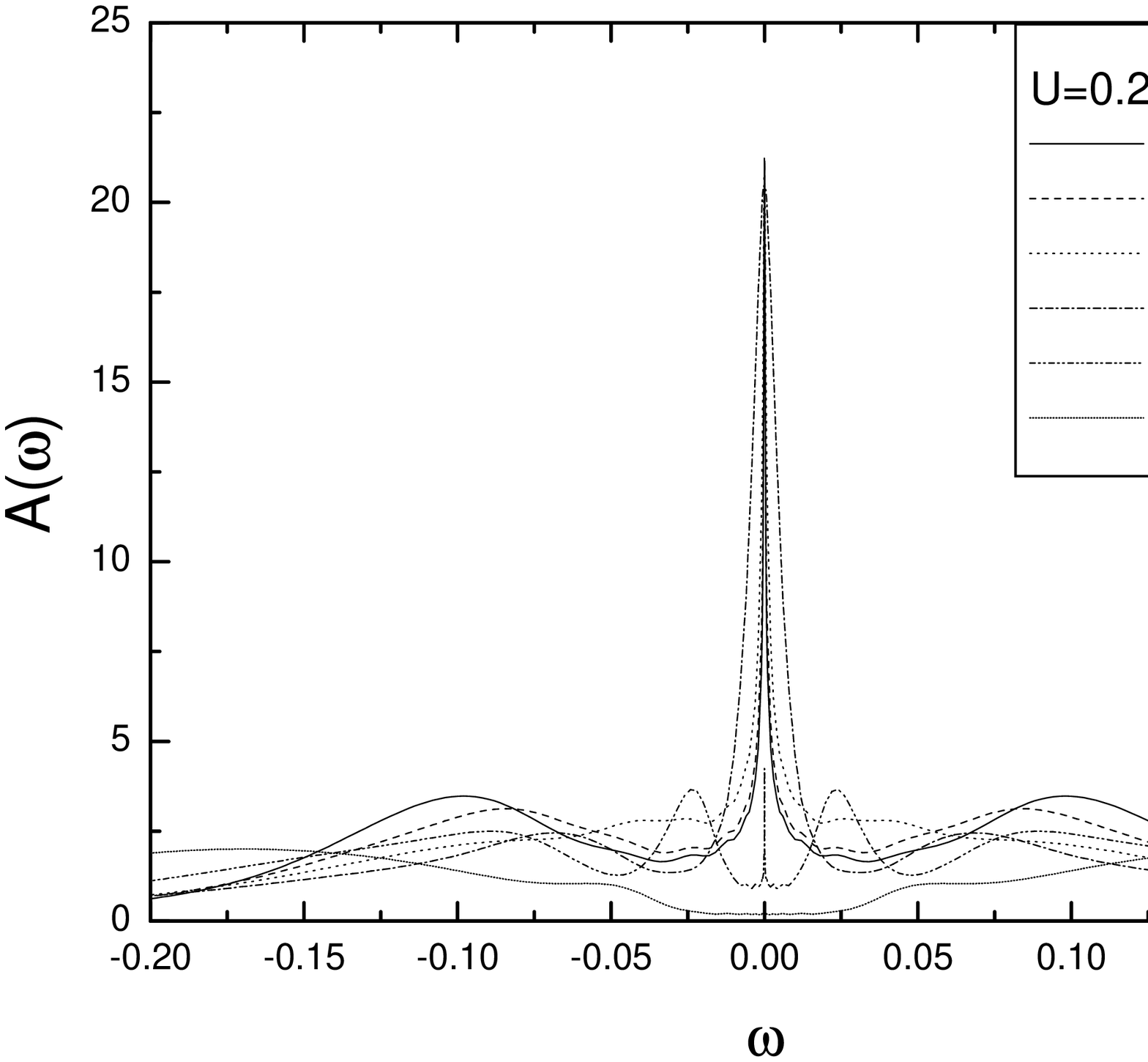}
\vspace{-5cm} \caption{The phonon and electron spectral functions
vs.\ $\omega$ for $U=0.2$ and $\Delta=0.016$. }
\end{figure}

\begin{figure}
 \label{fig5}
\vspace{5cm}
\includegraphics[scale=0.5]{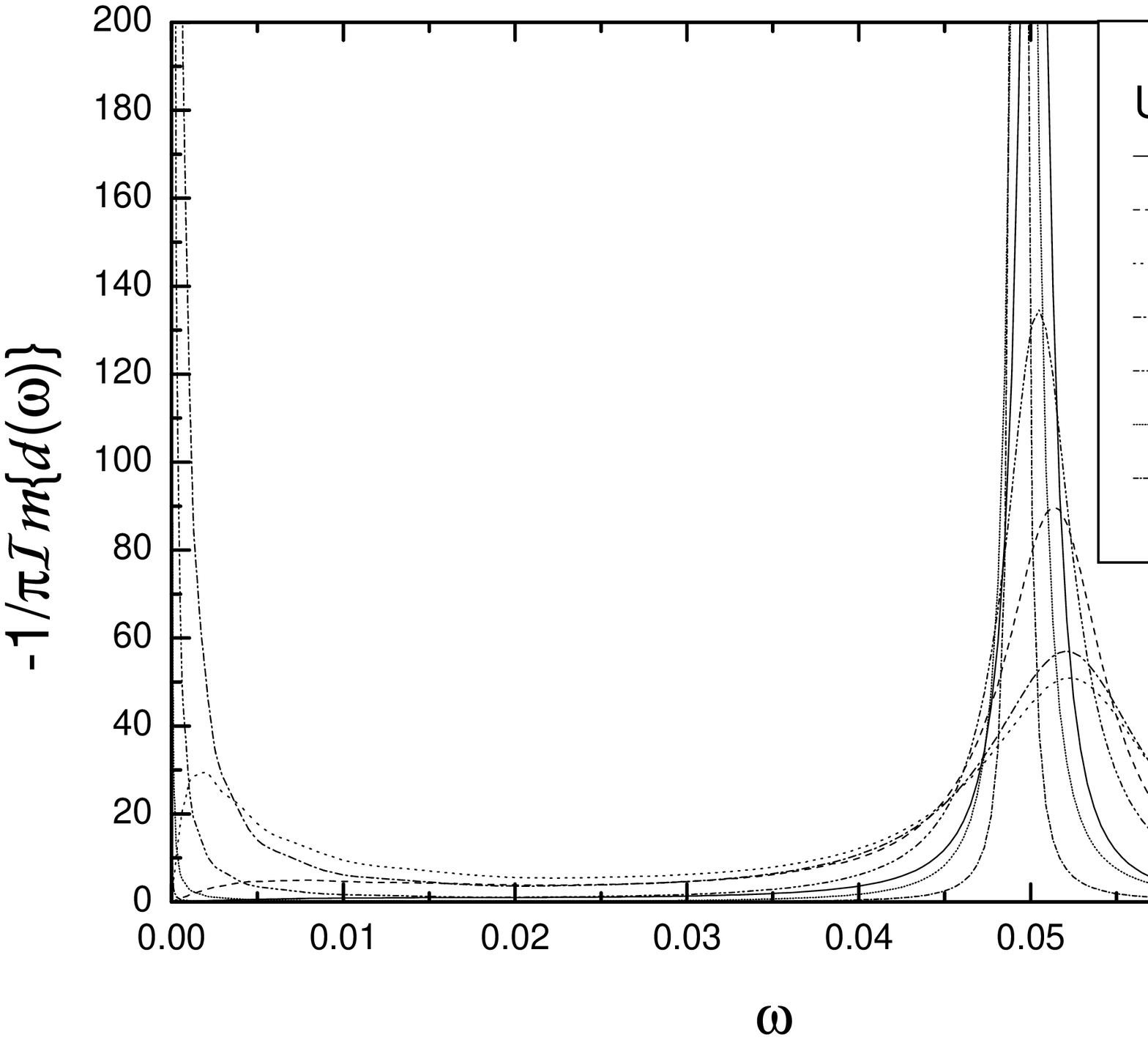}
\includegraphics[scale=0.5]{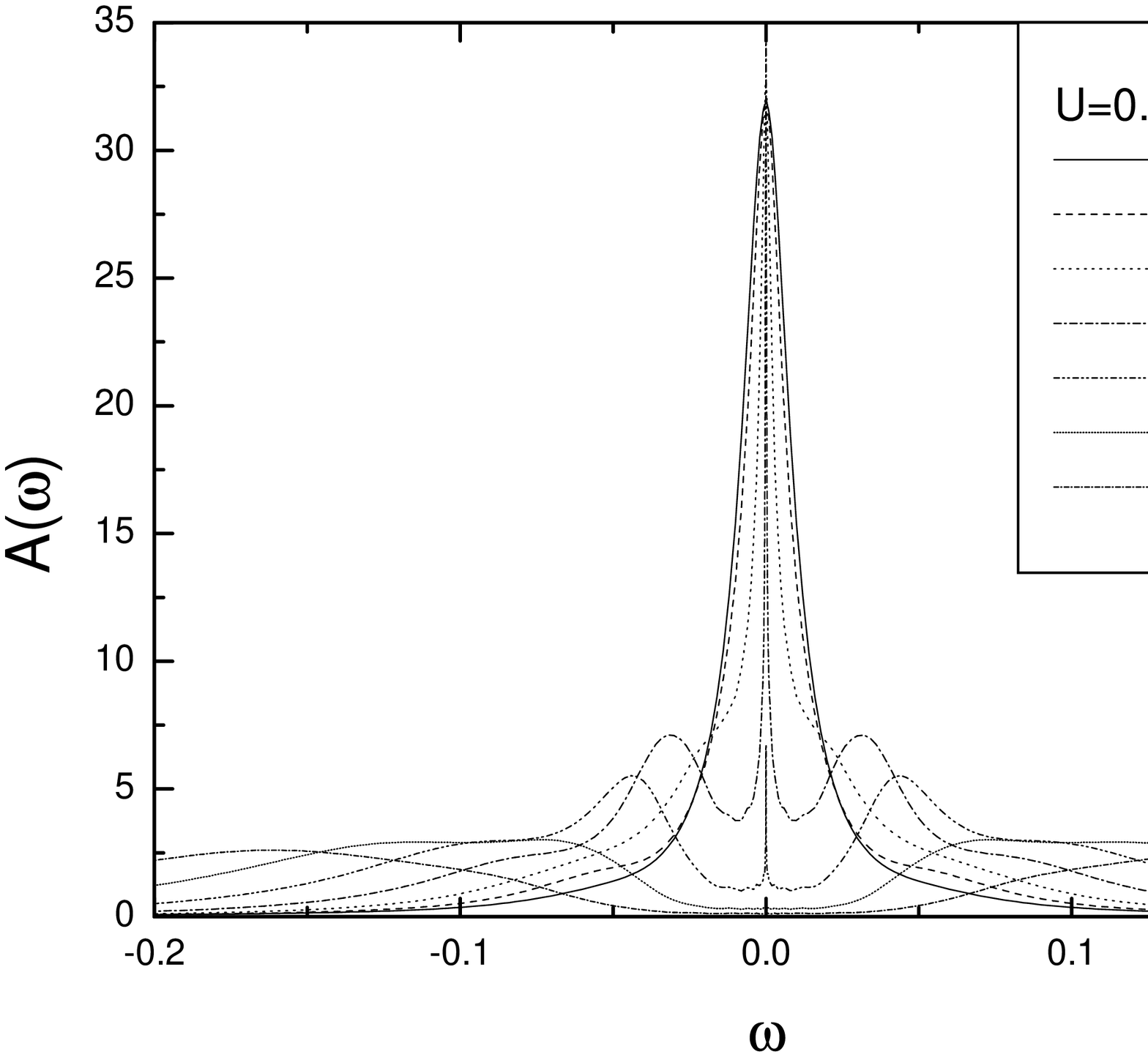}
\vspace{-5cm} \caption{The phonon and electron spectral functions
vs.\ $\omega$ for $U=0$ and $\Delta=0.01$. }
\end{figure}

We now present the results calculated with the method (b) for
other parameter values. Fig.\ 4 is for $U=0.2$ and $\Delta=0.016
\approx \omega_0/\pi$, and Fig.\ 5 is for $U=0$ and
$\Delta=0.01$. They exhibit the similar behaviors as for the
$U=0.1$ and $\Delta=\omega_0/\pi$ of figures 2 and 3. The
emergence of the soft phonon modes and suppression of the central
peak are correlated for both cases and agree with the estimate of
the perturbation theory of Eq.\ (\ref{soft}) as discussed above.
$g_{co} = 0.0613$ for $U=0.2$ and $\Delta=0.016$ and $g_{co} =
0.0280$ for $U=0$ and $\Delta=0.01$ which are consistent with the
NRG results of figures 4 and 5, respectively. The four high energy
peaks for negative $\bar U$ are also consistent with the small
polaron estimate of Eq.\ (\ref{polaronpeaks}). For $U=0.2$, shown
in Fig.\ 4, $\epsilon_f = -0.1$, and $E_p = 0.128$ for $g=0.08$.
We therefore expect to see the peaks in the electron spectral
function at $\omega \approx \pm (E_p +\epsilon_f),~\pm (E_p
+\epsilon_f +\omega_0),\cdots \approx \pm 0.028$, $\pm 0.078$, and
so forth, which agree with Fig.\ 4. For $U=0$, shown in Fig.\ 5,
$\epsilon_f =0$, and $E_p = 0.032$ and 0.05 for $g=0.04$ and 0.05,
respectively. The peaks are expected at $\omega \approx \pm
0.032$ and $\pm 0.082$ for $g=0.04$, and at $\omega \approx \pm
0.05$ and $\pm 0.1$ for $g=0.05$, which are again consistent with
the electron spectral function $A(\omega)$ shown in Fig.\ 5. The
heights of the central peak of the electron spectral function
$A(\omega)$ are given by $1/(\pi\Delta)$ of Eq.\ (\ref{height})
for both cases.

For $U=0$, the phonon renormalization seems somewhat different
from the $U\ne 0$ cases of Figs.\ 2 and 4 in that the main peak
around $\omega_0$ is slightly hardened as $g$ is increased. This
apparent difference can be straightforwardly understood from the
Green's function perturbation results presented in the previous
section. If one solves for the renormalized phonon frequency from
$D^{-1} (\omega) = 0$ using Eqs.\ (\ref{ren-d}) and (\ref{chi}),
one can see that the renormalized frequency gets softened or
hardened depending on $\Delta$ and $U$. The value of $U_s$ above
(below) which the renormalized phonon frequency, as $g$ in
increased from 0, is softened (hardened) is simply given by
$Re[\chi(\omega_0)/(1-(U/2)\chi(\omega_0))]=0$ from Eq.\
(\ref{ren-d}), which yields
 \ba
 \label{us}
U_s (\Delta) = 2\frac{Re\chi(\omega_0)}{|\chi(\omega_0)|^2} =
\frac{2\pi\omega_0^2}{\Delta}\, \frac{\ln\left(1+ \frac{
\omega_0^2}{\Delta^2} \right) -4 \left( \frac{ \Delta}{\omega_0}
\right) \tan^{-1}\left( \frac{\omega_0}{\Delta} \right)} {\ln^2
\left(1+ \frac{\omega_0^2}{\Delta^2} \right) +
4\tan^{-2}\left(\frac{\omega_0}{\Delta}\right)}.
 \ea
For $\omega_0=0.05$, one finds that $U_s = 0.0636$ for
$\Delta=\omega_0/\pi$, and 0.1868 for $\Delta=0.01$. We therefore
expect that the phonon renormalization will get softened for
$U=0.1$ and $\Delta=\omega_0/\pi$, and $U=0.2$ and $\Delta=0.016
\approx \omega_0/\pi$, corresponding, respectively, to Fig.\ 2(b)
and Fig. 4, but hardened for $U=0$ and $\Delta=0.01$
corresponding to Fig.\ 5, in agreement with the NRG results
presented in those figures.

Note, however, that the phonon spectral function of $U=0$ has a
long tail towards $\omega=0$ compared with the $U\ne 0$ cases and,
therefore, the average phonon frequency gets softened as $g$ is
increased. When $g \gtrsim g_{co}$, the soft mode emerges in
addition to the peak around $\omega_0$, and the average phonon
frequency hardens back to the bare frequency of $\omega_0$ as $g$
is further increased above $g_{co}$ like the $U\ne 0$ cases. Also
noticeable is that the hardening or softening of the renormalized
phonon frequency for $U=0$ does depend on $\Delta$. From Eq.\
(\ref{us}), we expect that the renormalized frequency will get
hardened (softened) for $\Delta < (>)$ $0.410\ \omega_0 = 0.0205$.
The NRG results presented in Fig.\ 5 for $U=0$ and $\Delta=0.01$
are consistent with the estimate.

\section{summary and concluding remarks}

In the present paper, we have presented Wilson's numerical
renormalization group calculations of the symmetric
Anderson-Holstein model. The phonon and electron spectral
functions were computed as the electron-electron interaction $U$
and electron-phonon coupling $g$ were varied. The electron
spectral functions are in good agreement with the Hewson and Meyer
results. On the other hand, an improved method was presented for
calculating the phonon spectral functions and the results were
compared with Hewson and Meyer's. Unlike the method (c) of Eq.\
(\ref{def3}) where the phonon frequency is always tied to the bare
frequency, the present calculations show that as $g$ is
increased, the phonon mode is renormalized until the soft mode
emerges around $g \approx g_{co}$, and as $g$ is further
increased, the phonon mode splits into two components, one of
which approaches back to the bare frequency and the other
develops into the soft mode. We argue that the above behavior of
the phonon renormalization is physically more sensible and is a
correct picture in that (a) in the present method the phonon
renormalization is explicitly computed, and (b) the present
method satisfies the boson sum rule better than the previous
methods as listed in Table 1. These NRG calculations were also
compared with the standard Green's function results for the weak
coupling regime to understand the phonon renormalization and soft
mode.

Now that the calculation of the dynamical properties of the
impurity Anderson-Holstein model is successfully carried out, we
plan to extend these calculations to the lattice model of the
Hubbard-Holstein model within the framework of the dynamical
mean-field theory. These calculations will be relevant to
high-$T_c$ superconductors where recent photoemission experiments
give evidence of strong electron-phonon coupling~\cite{Lea01} and
to the manganites~\cite{Ram97}. Extension to the finite
temperatures for the impurity model and for the lattice model
should also be fruitful.

We would like to thank Hyun Cheol Lee, Tae Suk Kim, Ralf Bulla,
Dietrich Meyer, and Alex Hewson for helpful comments and
discussions. This work was supported by the Korea Science \&
Engineering Foundation (KOSEF) through grant No.\
R01-1999-000-00031-0 and Center for Nanotubes and Nanoscale
Composites (CNNC), and by the Ministry of Education through BK21
SNU-SKKU Physics program.

\end{document}